\documentclass[conference]{IEEEtran}
\IEEEoverridecommandlockouts
\usepackage{cite}
\usepackage{amsmath,amssymb,amsfonts}
\usepackage{algorithmic}
\usepackage{graphicx}
\usepackage{textcomp}
\def\BibTeX{{\rm B\kern-.05em{\sc i\kern-.025em b}\kern-.08em
    T\kern-.1667em\lower.7ex\hbox{E}\kern-.125emX}}
\begin{document}

\title{Certified Ethical Hacker v.10 Online Course\\
        a Case Study
}

\author{\IEEEauthorblockN{Tam N. Nguyen}
\textit{North Carolina State University}\\
tam.nguyen@ncsu.edu \\
https://www.linkedin.com/in/tamcs/}

\maketitle

\begin{abstract}
CEH v.10 Certification Self-study Course is an online course preparing learners for one of the most prestige cyber security certifications in the world - the Certified Ethical Hacker (CEH) v.10 Certification. Due to a pay wall and the practical rather than theoretical nature, most researchers have limited exposure to this course. For the first time, this paper will analyze the course's instructional design based on the highest national standards and related peer-reviewed published research works. The sole intention is to push the course to a higher ground, making it the best online course for cyber security. More importantly, the paper's instructional design evaluation strategy can well be extended and applied to any other online course' instructional design review and/or evaluation process.
\end{abstract}

\section{Introduction}
Cybersecurity is a very challenging field due to the fast-pace attack-defense chess game. With the threat landscape constantly evolves, cyber security awareness education becomes a crucial part in the sustainability and growth of any corporation. The Certified Ethical Hacker (CEH) v10 Online Self-study course \footnote{https://www.eccouncil.org/programs/certified-ethical-hacker-ceh/} is one of the shining stars, as professionals with the CEH certification are well sought after by many of the Fortune 500 companies. With the intention to help build this course to be the best of its kind, the paper performs an initial independent instructional analysis of the course, providing recommendations for future improvements. 

The paper's structure is as followed. Section 2 provides a brief overview of the course. Due to copyright concerns, the paper will not provide detailed screenshots of the actual course interface but rather, descriptions and basic component mapping. Section 3 focuses on pedagogical analysis. Great care was shown by employing a careful blend of the highest national standards for online course evaluation. Through five pedagogical evaluation categories of "Meaningful", "Engaging", "Measurable", "Accessible", and "Scalable", the section presents the course's strong points as well as potential areas for improvements. Section 4 focuses on the digital technologies behind the course's cyber range (lab). Section 5 offers potential remedies to the issues at hand with specific actionable details.

The main contributions of the paper include: a 30-point check sheet for evaluating the CEH v10 course, 11 identified national standard points that were not met, and five general directions for future improvements. As far as the author's knowledge, this paper is the first to analyze a prestige online cyber security program like the CEH v10 Online Preparation Course. Its contributions can be well extended to other cyber security related courses or be used for any other initial instructional design development/evaluation projects.

\section{Overview of CEH v.10\\Official Preparation Course}

CEH v.10 Online Self-study Course was designed to help potential test takers prepare for the CEH v.10 certification test. CEH certification is a well-recognized industry standard and is a core sought after certification by many of the Fortune 500 organizations. It is ANSI 17024 compliant covering latest topics that a practical cyber security consultant should know.

The course is in its 10th iteration while still sticking to its original ultimate goal - teaching cyber security professionals to think like a hacker in order to defend against hackers. The course covers five common phases of an attack cycle which are Reconnaissance, Gaining Access, Enumeration, Maintaining Access, and Cleaning up traces. The target audience are system administrators, network administrators and engineers, Web managers, Auditors, ethical hackers and other types of cyber security professionals.

CEH v10 course qualifies for the "Massive" and "Online" criteria because the course is 100\% online and its learners come from around the world \cite{GlobalKnowledgeTrainingLLC20162016Knowledge}. However, it is not qualified for "Open" because of a significant pay wall (\$1,899 for a package of education suite and certification exam voucher). This pay wall may contributes to the difference in learners motivations and behaviors but investigation into that hypothesis is beyond the scope of this paper. Therefore, it is assumed that the CEH v10 course can be held to the same standards as a regular MOOC (Massive Open Online Course).

Learning suite includes e-book; online learning management system with video lectures, online textbook, a note-taking app; and online labs. Labs can be launched from web browsers, allowing learners access to lab environments with latest operating systems, virtual network appliances, preconfigured enterprise systems, and so on. It is advised that learners spend more than 40\% on the virtual labs. There are 20 modules and after each module, there is a hacking challenge to help learners transform knowledge into skills that can be applied to real life situations. The engine behind the labs is actually the well known "Learn On Demand System" (LODS) which will be discussed in later section.

\begin{figure}[h]
  \centering
  \includegraphics[width=5cm]{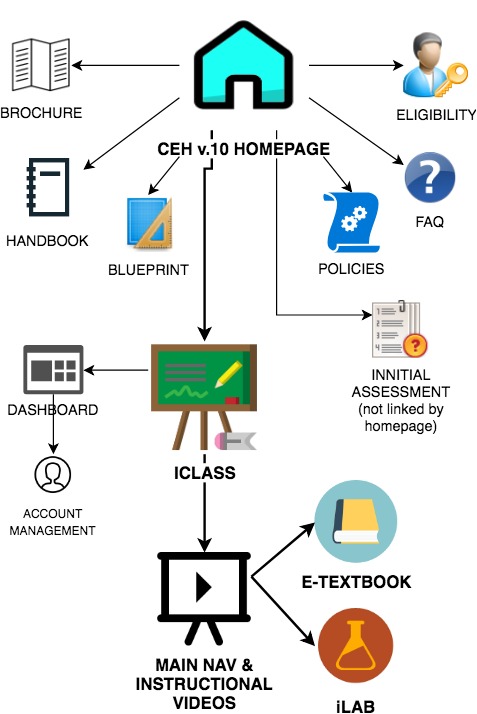}
  \caption{CEH v.10 Site Structure}
  \label{Figure:siteStructure}
\end{figure}

\section{Pedagogical Evaluations}

The paper aims to perform instructional analysis of the CEH v10 course from an outsider, "black-box" perspective with no insider's help regarding collected usage data or any other form of internal insights. The evaluation approach is a combination of actual course studying by a real learner with real intention to finish the course in order to take the certification exam, well-established standards, and previous research works on evaluation criteria for MOOC (Massive Open Online Course).

Because instructional design choices have "a significant impact on deep and meaningful learning" \cite{Garrison2005FacilitatingEnough} there are numerous studies on how to evaluate the designs of MOOC. Conole \cite{Conole2014AMOOCs} suggested MOOCs to be evaluated by twelve dimensions of: Open, Massive, User of media, Degree of communication, Degree of collaboration, Learning pathway, Quality assurance, Amount of reflection, Certification, Formal learning, Autonomy and Diversity.

QualityMatters \cite{QualityMattersHigherMatters} grouped evaluation criteria into eight groups of (i) Course overview and introduction (ii) Learning objectives (iii). Assessment and measurement (iv) Instructional materials (v) Learner interaction and engagement (vi) Course technology (vii) Learner support (vii) Accessibility (Quality Matters, 2014). Other evaluation methods include:  the Rubric for Online Instruction \cite{CaliforniaStateUniversity2009RubricInstruction}, iNACOL National Standards for Quality Online Courses \cite{INACOL2011National2}, and the five pillars of quality framework for quality online course design by the Online Learning Consortium \footnote{onlinelearningconsortium.org/about/quality-framework-five-pillars}.

In another attempt, Drake studied different evaluation methods as well as case studies and compressed everything down to just five principles for MOOC success \cite{Drake2015FiveStudy} which are Meaningful, Engaging, Measurable, Accessible and Scalable. This approach was embraced by the University of North Carolina (UNC) System in their request for proposals to develop a MOOC on Emerging Economies. Those five principles were chosen as the five objectives as detailed by Dr. Tom Ross, President of the UNC System \cite{Ross2014UNCPartnership}.

Because it is not common for MOOCs to satisfy the strict QualityMatters criterias for higher-education online learning course standards \cite{Lowenthal2015InMOOCs}, the paper takes a mixed approach of mapping the above-mentioned QualityMatters standards into the five success principles for online course by Drake, enriched with additional details from iNACOL National Standards.

The paper believes a group of five principles is the right amount at the right abstraction level for an initial informal online course evaluation. It is also more approachable to readers who do not have in-depth knowledge about instructional design and/or do not care much about fine-grained evaluation criteria. The side effect is the relaxation or even elimination of some QualityMatters check points. However, since most MOOCs do not meet those points \cite{Lowenthal2015InMOOCs} and with additional points from iNACOL standards, the paper does not give CEH v10 an unfair evaluation.

Pedagogical evaluation results are shown in Table \ref{table:evaluation}. 

\begin{table*}[!t]
\centering
\begin{tabular}{|l|l|c|c|}
\hline
\multicolumn{1}{|c|}{\textbf{OBJECTIVES}} & \multicolumn{1}{c|}{\textbf{CHECK POINTS}} & \textbf{MET} & \textbf{\begin{tabular}[c]{@{}c@{}}NOT\\ MET\end{tabular}} \\ \hline
\textbf{1. Meaningful} & \begin{tabular}[c]{@{}l@{}}+ Introductory materials gives learners a good picture of course layouts,\\ order and schedules as well as navigational instructions (QM 1.1)\end{tabular} & x &  \\ \hline
 & \begin{tabular}[c]{@{}l@{}}+ Introductory materials presented clear purpose of the course and how \\ the learning process will be (QM 1.2)\end{tabular} & x &  \\ \cline{2-4} 
 & + Prerequisite and/or competencies are well communicated (QM 1.7) &  & x \\ \cline{2-4} 
 & \begin{tabular}[c]{@{}l@{}}+ Learning objectives and expected learning outcomes  is presented in a\\ way that is easy to understand to all learners, including the ones with\\ difficulties (QM 2.3)\end{tabular} & x &  \\ \cline{2-4} 
 & \begin{tabular}[c]{@{}l@{}}+ Learning activities are well explained as to how accomplishing those\\ activities will help the learners reach the planned learning competencies\\ (QM 2.4)\end{tabular} & x &  \\ \cline{2-4} 
 & \begin{tabular}[c]{@{}l@{}}+ Levels of learning are well organized and match the expected content\\ mastery (QM 2.5)\end{tabular} & x &  \\ \cline{2-4} 
 & + Instructional materials are up-to-date (QM 4.4) & x &  \\ \hline
\textbf{2. Engaging} & + Communication netiquette is well communicated (QM 1.3) &  & x \\ \hline
 & + Communication plans are clearly stated (QM 5.3, 5.4) &  & x \\ \cline{2-4} 
 & \begin{tabular}[c]{@{}l@{}}+ Instructor was able to create a sense of connection with the learners,\\ being approachable (QM 1.8)\end{tabular} & x &  \\ \cline{2-4} 
 & \begin{tabular}[c]{@{}l@{}}+ Introductory activities help create a welcoming learning environment, \\ and a sense of community (QM 1.9)\end{tabular} &  & x \\ \cline{2-4} 
 & + Learners' sense of achievement is frequently promoted (QM 5.1) &  & x \\ \cline{2-4} 
 & \begin{tabular}[c]{@{}l@{}}+ Interactive activities and Active learning is encouraged through\\ meaningful interactions that align with the course objectives (QM 5.2)\end{tabular} & x &  \\ \cline{2-4} 
 & + Course tools promote active learning and engagements (QM 6.2) & x &  \\ \cline{2-4} 
 & \begin{tabular}[c]{@{}l@{}}+ Feedbacks are timely, accurate and in various forms coming from both\\ instructors, peers and tools (QM 3.5)\end{tabular} &  & x \\ \hline
\textbf{3. Measurable} & \begin{tabular}[c]{@{}l@{}}+ Learning competencies/objectives were described clearly using terms\\ that are specific, measurable, and observable (QM 2.1, 2.2)\end{tabular} & x &  \\ \hline
& \begin{tabular}[c]{@{}l@{}}+ There are assessments to measure the stated learning competencies\\ (QM 3.1)\end{tabular} &  & x \\ \cline{2-4} 
 & \begin{tabular}[c]{@{}l@{}}+ Learners are provided multiple ways to demonstrate progress and\\ mastery of the competencies (QM 3.4)\end{tabular} &  & x \\ \cline{2-4} 
 & \begin{tabular}[c]{@{}l@{}}+ Progress tracking mechanisms are provided to both instructors and\\ learners (QM 3.5)\end{tabular} & x &  \\ \hline
\textbf{4. Accessible} & \begin{tabular}[c]{@{}l@{}}+ Technology requirements as well as needed technical skills were\\ clearly communicated prior to the course (QM 1.5, 1.6)\end{tabular} &  & x \\ \hline
 & \begin{tabular}[c]{@{}l@{}}+ A variety of instructional materials is used, contributing to the stated\\ learning objectives, containing well-referenced sources (QM 4.1, 4.3,\\ 4.5)\end{tabular} & x &  \\ \cline{2-4} 
 & \begin{tabular}[c]{@{}l@{}}+ Tools used are of various types, easily accessible, relevant to course\\ objectives while protecting learners' privacy (QM 6.1, 6.3, 6.4)\end{tabular} & x &  \\ \cline{2-4} 
 & \begin{tabular}[c]{@{}l@{}}+ Instructions, Privacy policies, Accessibility policies and other Support\\ documents are clearly accessible to learners (QM 6.4, 7.1, 7.2, 7.3, 7.4)\end{tabular} &  & x \\ \cline{2-4} 
 & \begin{tabular}[c]{@{}l@{}}+ Course navigation system was well-designed and intuitive to learners,\\ reflecting thoughtful strategies that promote effective learning (QM 8.1)\end{tabular} & x &  \\ \cline{2-4} 
 & + Course designs maximize usability and efficient learning (QM 8.2) & x &  \\ \cline{2-4} 
 & \begin{tabular}[c]{@{}l@{}}+ Course designs bear no barrier to learners with disabilities, accessible\\ design practices are followed (QM 8.3)\end{tabular} & x &  \\ \cline{2-4} 
 & \begin{tabular}[c]{@{}l@{}}+ Course multimedia facilitate ease of use with alternative means of\\ access (QM 8.4, 8.5)\end{tabular} &  & x \\ \hline
\textbf{5. Scalable} & \begin{tabular}[c]{@{}l@{}}+ The course is designed to meet internationally recognized\\ interoperability standards. (iNACOL)\end{tabular} & x &  \\ \hline
& \begin{tabular}[c]{@{}l@{}}+ Copyright and licensing status, including permission to share where\\ applicable, is clearly stated and easily found. (iNACOL)\end{tabular} & x &  \\ \cline{2-4} 
 & \begin{tabular}[c]{@{}l@{}}+ The course accommodates multiple school calendars, schedules, etc.\\ (iNACOL)\end{tabular} & x &  \\ \hline
 \\
\end{tabular}
\caption{CEH v10 Evaluation Check Points}
  \label{table:evaluation}
\end{table*}

\subsection{Meaningful}
Starting with course introduction, learners should be able to quickly recognize what is this course about, the order of learning objectives, schedules and how to generally navigate the course. Prerequisites and/or competencies are well communicated with easy to find links. Description of expected learning outcomes is presented in a way that is easy to understand even to learners with difficulties/disabilities. Learning activities are well-explained as to how accomplishing those activities will help learners acquire the necessary skills.  The levels of learning and the relationships between course components within each level are well organized and match the expected content mastery for each level. Instructional materials are all up to date.

Good signs include but are not limited to cognitive and meta-cognitive prompts, short distilled lectures on single topics, study guides, concept maps, self-assessment quizzes, discussion board. Bad signs may include irrelevant topics, poor idea integration, confusing order, insufficient examples.

Evaluation shows that the course's introductory materials provided really good details on the course layout mostly in the form of "Table of Content". Because this is a self-paced course, no expected schedule was listed. However, it was emphasized that learners will have up to 1 year of access to the instructional materials and 6 months of access to the lab. Explicit navigational instruction was not listed but rather enforced by noticeable visual clues like large buttons. The purpose of the course and how the learning process should be followed are apparent.

The course's module objectives are listed at the beginning of each module and there is a summary at the corresponding end. Learning activities which are reading text book and doing labs were emphasized at during course orientation. Levels of competencies are well organized and instructional materials are all up to date.

Prerequisite for the course were not well communicated. There is an assessment test \footnote{eccouncil.org/programs/certified-ethical-hacker-ceh/ceh-assessment/} but it stays completely separate from the official training site \footnote{https://iclass.eccouncil.org/learning-options/}. There is also no dedicated page on required tech skills, hardware and software for the course. However, there is always an option to chat or leave a message with a representative at the bottom right of each page in the main e-course interface.

\begin{figure}[h]
  \centering
  \includegraphics[width=6.5cm]{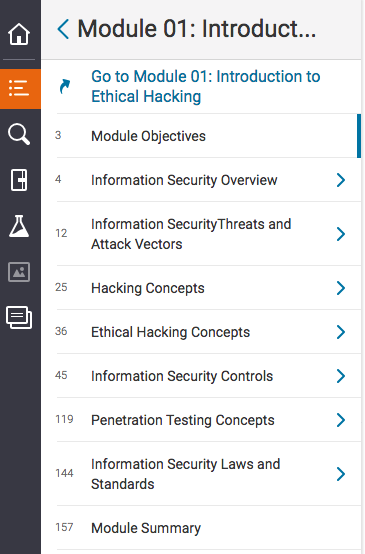}
  \caption{E-textbook Core Navigational System}
  \label{Figure:Core Navigational System}
\end{figure}

\subsection{Engaging}

Since it is very common to have just 5\% of registered learners finished a course \cite{Daniel2012MakingPossibility}, we should engage learners as often as possible. There are Cognitive engagement (task-specific thinking), Emotional engagement (affective responses towards learning, teachers and peers) and Behavioral engagement (participation in an activity that leads to a completion of something) driven by the psychological needs of autonomy, relatedness and competence \cite{Hew2016PromotingMOOCS}. It is also important to note that there are different groups of students with different interests. There are students who valuate the importance of course atmosphere, the importance of exercise, the importance of teacher, or the importance of exams more than the others \cite{Holstein2016ThePerception}.

An engnaging course starts with a clear guideline on communication netiquette, and clear communication plans. Since participants of a course come from different regions of the world, it is important to have guidelines that embrace and respect all cultures. Instructors are able to establish a sense of connection with the learners right from the beginning by professional in content, friendly in tone, proper visual appeals and so on. Learners, on the other hand, should also be given the opportunity to introduce themselves, promoting a sense of community.

Active learning such as discovering, processing, or applying concepts and information is encouraged through meaningful activities, interactions and tools that align well with course objectives as well as expected levels of mastery. Through learner-instructor interactions (discussion board exchanges, FAQ,...), learner-content interactions (assigned readings, assigned workbook, online exercises,...), learner-learner interaction (group discussions, small-group projects, group discussions, peer critiques,...), learners' sense of achievement should be frequently promoted.

Good signs include but are not limited to short videos, bite-size assignments with immediate feedback (automated grading for example, automated emails, automated reminders), discussion groups, virtual chat rooms, even local meet-ups \cite{Sugar2010ExaminingStrategies}.

The course evaluations shows that the instructor has decades of teaching IT courses and was able to establish trust, coming across as an approachable professional. The major and probably the only force that drives engagement is the iLabs which basically provides learners a virtual environment with real solution stacks to exercise and experiment with.

What missing is a sense of community with no discussion board within the online course itself. There is no way to communicate with other self-pace learners, no quick way to reach the instructor or mentors. Consequently, there is no established guideline on communication netiquette, no communication plan. Feedback from instructor and peers are unavailable.

The course is also lacking in building methods to constantly reinforce learners' sense of achievement. Except for clicking the "complete" check box after each module, learners receive no other success indicator or encouragement from the system. Together with the lack of a built-in discussion board may make the learner feel really lonely in this challenging course.

\subsection{Measurable}
On-line courses should be measurable from both of the course designers' and the learners' perspectives. Learning competencies and objectives should be described clearly using terms that are simple, specific, measurable and observable. One example is a short description of what learners will learn and what learners should be able to perform after each module. At the same time, there are measures to immediately identify if competencies are met. Because learners come from a broad and diverse background, assessments should be provided in various forms and formats to give learners multiple ways to demonstrate their mastery of corresponding competencies. Examples of various assessments include polls, one question quizzes, quizzes, short essays, short programming, reaction videos, forum posts, etc.

Progress tracking mechanisms should be provided to both instructors and learners. Examples include: voting buttons, completion check boxes, self-mastery tests, interactive simulations, self-scoring quizzes, automatic grading of programming assignments, practice written assignments that receive feedback, peer review papers,etc.

The course's learning competencies and objectives were described clearly using terms that are specific, measurable, and observable. Progress tracking mechanisms are clearly visible. While there are multiple ways for learners to demonstrate progress, there is only one way for learners to demonstrate the mastery of the competencies - finishing the iLab challenges.

There are at least two main big disadvantages of using just iLab challenges as the only true measurement of learners' knowledge. First, the iLabs only appear at the end of each module. Therefore, it is not possible to measure learners' comprehensions of the sub-modules prior to the iLab sub-modules. Second, due to the nature of a real but virtualized environment, it is impossible to link certain mistakes made in the iLab environment with corresponding instructional section for later reference. 

Due to those reasons, besides the course progress, learners do not have access to further statistics regarding their mastery levels. For example, learners do not know at a particular point in time what are the learning points that they are most good or bad at. 

\subsection{Accessible}

Required technologies as well as technical skills were clearly communicated prior to the course such as needed webcam, plugins, mobile applications, etc. A variety of instructional materials is used, being aligned with the stated learning objectives, containing well-referenced resources. Instructions, privacy policies, accessibility policies and other support documents are clearly accessible to learners.

Course navigation system was well-designed and intuitive to learners, reflecting thoughtful strategies that promote effective learning. Course designs maximize usability and bear no barrier to learners with disabilities. Accessible design practices are followed. Course multimedia facilitate ease of use with alternative means of access.

A variety of instructional materials is used including videos, e-text book, simulated environments, text-to-speech, flash cards,etc. Tools used are of various types, easily accessible and relevant to stated course objectives. Course designs maximize usability and efficient learning by limiting distractions, driving the focus to what really matters. The course also has a good navigation system including hierarchical index tree, bookmarks, slide bar for navigation, search bar, figure browsing, and in-page notebook/scratchpad. Accessibility functions are decent and include "Read aloud", "Night display" and captions for instructional videos.

Areas of improvements for this section include: a more accessible privacy policies, accessibility policies, instructions and self-support documentations; more accessible instructional videos that cater to users who are not fluent in English, have limited bandwidth, users with hearing difficulties.

\subsection{Scalable}
This principle deals with how the architecture, funding, and content development of the course allow it to be scaled. The course design has to meet internationally recognized interoperability standards, supporting multiple school calendars, schedules. Copyright and licensing status must be clearly stated.

Scalability is probably the strongest point of this online course. With decades of training learners internationally, the designers of CEH v10 know how to design an educational system that is easy to update and scalable. At the moment, they are still improving and expanding the course. Further details regarding scalable technologies will be discussed in section 4.

\section{Cyber Range Evaluation}
CEH OPC lab environment is powered by "Learn On Demand Systems" (LODS) of which history can be dated back 25 years. LODS has extensive experience in providing simulation platforms for technical training, creating training contents with clients from all over the world including certification authorities like CompTIA, EC-Council, ISC2 and big international corporations like Google, Microsoft. At the moment, there are 13,500,000 labs launched; 40,000,000 VMs deployed; 5,000,000 students trained with LODS \cite{LODSLearnSolutions}. This section highlights several key features that differentiate LODS from other cyber simulation systems.

\subsection{Flexible and scalable virtualization technologies}
LODS supports multiple virtualization strategies and platforms, including Azure \cite{Azurecat2018Structured2}, Amazon \cite{Mathew2014OverviewServices}, Hyper-V \cite{2013WhyV1.0} and VMware \cite{VMwareInc.2007VMwareOverview}. The system has a capability portfolio of hosting in Local Datacenters, hosting on a Cloud Platform, testing of virtualizations, automatic Screen Scaling, easy migration and moving of Virtual Machine (VM) images and disks, highly configurable Networking, Advanced Network Interfacing/Monitoring, and high User Concurrency supporting over 5000 users simultaneously from a single location.

Its "Cloud Slice" technology offers the flexibility of integrating directly with one's own cloud provider on parts of the overall training system, in conjuntion with LODS clouds, with independent virtual machines hosted on bare-metals, and even with evironments that do not use virtual machines. All VMs are accessible through web browsers without users' overhead of downloading and installing new client softwares.

Through IP tracking and API, LODS allows smart caching and load-balancing of labs. On one hand, geo-location information derived from clients IPs hints LODS on preferred geo-location. Lab files will then be proactively replicated to corresponding data-centers. On the other hand, if a region is being over-loaded with demands, users' requests can be forced-directed and be handled by data-centers from different regions.

\subsection{Activity-based Assessments}
Activity-based assessments include automated checks and quizzes as two main ways to check learners' knowledge while they are working on a lab. Quizzes have traditional options such as multiple choices, true or false, type-in answers to be checked by regex, etc. Automated checks are scripts configured to run against VMs and/or other virtual instances/appliances. 

These scripts look for forensic evidences of certain actions that are supposed to be performed by the learners. For example, automated scripts can check if learners have been following the recommended steps correctly. Automated scripts can also be used to help the learners recognize and understand their mistakes, as well as to give the confirmations and encouragements as needed. Finally, automated checks can also be leveraged to perform lab steps that are supposed to be performed by the system but only in response to certain actions performed by the learners. Scoring can be switched on or off for both automated checks and quizzes.

In addition, learners can manually initiate system checks by using the "On-demand evaluation" button whenever they are done with a particular task list. On the other hand, some tasks are required before learners may advance to the next step. Feedback will be displayed to the learners after their answer submissions.

\begin{figure}[h]
  \centering
  \includegraphics[width=7cm]{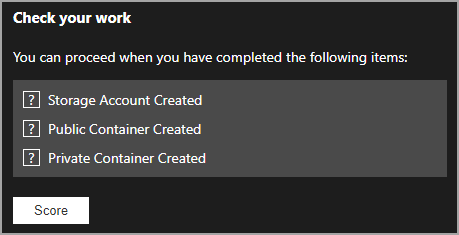}
  \caption{User view of tasks}
  \label{Figure:Tasks}
\end{figure}

\subsection{Shared-lab and role-playing}
Interestingly, LODS also supports the Shared-lab model with role-playing among learners with instructor's supervision and orchestration. This model mimics real-world situations where there are different teams working on the same infrastructure, supporting each other. Learners will be assigned different roles with different responsibilities, different tasks but they will all be working in the same lab environment. This model encourage them to cross-train, collaborate, communicate and learn from each other. In a broader scale, the model can be extended into massive cyber exercises where teams from different organizations, different geographical locations can participate at the same time. Finally, the shared-lab model helps with reducing operational costs.

\subsection{Integrated Life-cycle Actions}
Another great feature is the Integrated Life-cycle Actions where the lab can be integrated and/or dynamically extended to other platforms via main actions of sending web requests, sending emails and execution of commands. Whenever there is an event happening inside LODS, one or several associated actions can be triggered. The sending of web requests can be "GET" or "POST (inbound or outbound), url based with parameters corresponding to specified APIs of the destination applications. The sending of notifications can be sent via the lab interface or be sent via emails. The execution of commands requires learners to be logged on a VM and can provide powerful effects. All of these features technically allow LODS to be "pluggable" with all other systems, which may include Machine Learning/Artificial Intelligence As-A-Service systems, intelligent tutoring systems, etc.

Most importantly, LODs supports the Learning Tools Interoperability (LTI) standard \footnote{https://www.imsglobal.org/activity/learning-tools-interoperability} with main features of: Deep linking (more intuitive way to add and link contents from learning tools or publisher library), Assignment and Grade Services (exchanging assignemnt progresses and scores between platforms), Provisioning service (specifying roles and protecting security as well as privacy). For example, LODS can be integrated seamlessly with edX (a well-known MOOC platform) as a LTI provider. Progresses and scores can be instantly recorded by edX system for immediate analysis and appropriate actions.

\subsection{Lab Managements}
General management of labs include creation, design, import, export, cloning, etc. An administrator may delegate class editing to another user. Courses can be created from scratch with videos, virtual labs, assessments, surveys and other downloadable materials. However, courses can also be built from trusted content providers such as Microsoft, Logical Operations, EC-Council, GTSLearning, IBM and VMware. Automatic check-in/check-out will be done whenever a course is being edited in order to manage changes effectively. Course statistics can be easily exported securely. For example, course survey results may be exported via API using the ExportSurveyResponses method with the option to remove names and emails.

Powerful announcement feature allows course managers to make announcements and notifications around events, on demand subscriptions, enrollments and other related learning processes. Announcements may be prioritized into levels of mandatory, high, or normal, appearing in specific locations with mutable expiration dates. Announcements can also be time-based - a very useful feature considering time is on of the main interests for online courses. For example, a time-based reminder may be automatically sent to learners who may be late for an assignment submission.

\section{Recommendations}
In studying the successful recipes from three highly rated MOOCs \cite{Hew2016PromotingMOOCS}, Hew confirmed that the three psychological needs of Autonomy, Relatedness and Competence play critical roles in student online course engagements, which in turn are the key to the MOOCs' success. The factors of "peer interaction" and "instructor accessibility" were found to be half of the driving forces behind those important psychological needs. Therefore, the paper will center its recommendations around these two important factors.

\subsection{Create a stronger sense of community}
Peer-learning was found to be highly beneficial to cyber security education \cite{Pittman2015AnCamp}. A discussion board should be built into the online course and be used by self-paced learners to communicate with others including mentors. There can also be leader boards, hall of fame, or even a section to set up time and dates for local meet-up events. Limited amount of information regarding learners progresses can be made available to the community so that members can know where they are among others in term of the course's progresses and achievements. Certified learners can be invited to be mentors, helping instructors out with answering questions, giving tips, even sharing own struggles while studying the course. Research have shown that the act of tutoring others will boost learners' conceptual elaboration, enhancing the transferability of acquired knowledge \cite{Walker2008ToTutoring}.

There must be a strong guideline for community engagements, respecting differences among the diverse group of international learners. There should also be a communication plan covering as many possible situations as possible. For one example, after a learner has been in active for weeks, the system may contact the learner via mobile text messages, checking if he or she is having some issues with the course.

\subsection{Quizzes as a measurement tool}
Quizzes may appear to be insignificant in the context of an online preparation course preparing learners for the certification test rather than a GPA number. However, quizzes are more than just a quantitative tool - they can also be sensors used to evaluate meaningful learning \cite{Wei2017IntegratingAssessment}. Identified by Ausubel (Ausubel, 1963), meaningful learning happens within one's cognitive structure by substantively integrating new knowledge with existing relevant ideas. However, cognitive structure is not visible and has to be mapped out to the real world by methods such as Concept Mapping. Quizzes can help with building such Concept Maps (CMs) to be evaluated by both learners and the instructors. From such CMs, interventions can be executed to prevent permanent misconceptions. It is important to note that a learner with misconceptions may still be able to pass the certification test. 

The initial assessment test should be a part of the course introduction. In fact, the common practice among most certification preparation courses is having the learners do a full simulated test, with the same time length and the same amount of questions. After finishing the assessment, learners will be presented with a strength map, indicating the areas they are most good at and the areas they should pay more attentions to.

At the end of each sub-module, there should be a one-question quiz. A short quiz with three to five questions should be presented after the summary of each module. The quizzes together with the initial assessment test can be used to measure learners' mastery of the module objectives. On the course designers' side, such measures can be used to plan incremental course upgrades. This is especially important when the CEH v10 course is a living online course.

\subsection{More ways to show competencies}
At the moment and within the scope of the course's current version, the only way for learners to prove their mastery of a module's educational objectives is to successfully finish the iLab for that module. Access to the iLab is limited to 6 months which is half of the time for the entire course. While learners are provided options to buy and extend access to iLab and while it makes sense that the best way to test an ethical hacker is through actual labs, the course still need to add more ways for the diverse group of learners to show their content mastery. It is important to note that the iLab still requires learners to be in a very specific setting to carry on the labs.

Options may include short quizzes as previously mentioned, Q\&A, relevant news sharing, mobile messaging quizzes, polls, etc. Q\&A model allows a learner to act as a peer mentor and answer questions using his/her own interpretation of the acquired knowledge. Whenever learners spot news about hacking incidents, pieces of malware, or any potential emerging threats relating to what had been taught in the course, they can post the news to the board for further discussions. Mobile messaging quizzes gear toward learners with extremely low bandwidth or learners who are frequently on the go. A short True/False question will be sent to the learners and all they need to do is replying back with either "T" or "F". Polls can be a good way for learners to learn from community consensus about a highly debatable topic.

\subsection{Providing feedback to learners more often}
Learners should be encouraged more often. It can be done after each time they finish a quiz or an important milestone, finish a lab within a short amount of time, reaching a certain page in the e-textbook, or spending a good amount of study time for several days straight. Encouragements may come in a form of simple congratulations, non-monetary rewards, an interesting fact, or even a bonus hidden problem for learners to tackle, etc. Encouragements sent from the system are even more important when interactions with the instructors are limited. Encouragements help maintain affective engagement and even improve cognitive engagement \cite{Hew2016PromotingMOOCS}.

From reported behavior studies, affective feedback when done properly may alter learners' cyber-security related behaviors for the better \cite{ShepherdSecurityBehaviour}. This is particularly useful in cases where learners were unaware of their previous risky behaviors. This and other types of feedback - especially the one informing learners of potential misstep - could be provided by a low-cost, adaptive system \cite{FengAuthoringTutoring}.

\subsection{Make videos more accessible}
Instructional videos should be made more accessible to all types of learners. Video transcripts should be downloadable to help those with extremely limited bandwidth. Video player should have speed adjustment functions. Learners who are familiar with the competencies targeted by the videos may up the play speed. Learners who are not proficient with English may slow the videos down. Transcripts of videos should also contain links to related course materials to make it easier for learners to cross-reference key knowledge points.  

\section{Conclusions}
Overall, the CEH v10 Preparation Online Course was well designed but there are still plenty of rooms for improvements. The cyber threat landscape is constantly evolving and the designers of the course honestly admitted that the whole course is a living one with the course contents are still being developed, matured, and extended.

The paper embraces this approach and contributed a complete sheet of 30 checkpoints categorized into five instructional design objectives of: Meaningful, Engaging, Measurable, Accessible and Scalable. The current state of the course does not meet 11 check points which lead to the paper's second contribution of five main recommendations with a focus on building a better community for learners within the course. "Self-paced" does not mean "working alone". By building a community within this pay-walled course, more values can be offered to learners. Since it is community-based, the total investment costs can be really affordable. The paper hopes CEH v10 course designers will receive the recommendations with open arms and future improvements will arrive soon.

Finally, the paper hopes its strategy for evaluating an online course can be adopted by companies for their initial evaluations of the courses or the training platforms they will invest in. Smart measurement methods must be implemented in order for early detection of misconceptions, even minor ones. Smart measurement also means better statistics leading to better justifications of the quality the education system is providing.

\bibliographystyle{abbrv}
\bibliography{references.bib}

\end{document}